\documentclass[prb,showpacs,preprint,nofloats,endfloats]{revtex4}

\usepackage{amsmath,amssymb,epsfig,epsf,psfig}

\renewcommand{\vec}[1]{{\mathbf #1}}

\begin{document}

\title
{\bf Pseudogaps in Strongly Correlated Metals}
\author{M.V.Sadovskii$^1$, I.A.Nekrasov$^{1,2}$, E.Z.Kuchinskii$^1$,
Th.Pruschke$^3$, V.I.Anisimov$^2$}

\affiliation
{$^1$Institute for Electrophysics, Russian Academy of Sciences,
Ekaterinburg, 620016, Russia\\
$^2$Institute for Metal Physics, Russian Academy of Sciences, 
Ekaterinburg, 620219, Russia\\
$^3$Institut f\"ur Theoretische Physik, Universi\"at G\"ottingen, Germany
}

\begin{abstract}
We generalize the dynamical--mean field (DMFT) approximation by including
into the DMFT equations some length scale via a (momentum dependent) ``external'' 
self--energy $\Sigma_{\bf k}$.  This external self--energy describes non-local
dynamical correlations induced by short-ranged collective SDW--like antiferromagnetic spin (or 
CDW--like charge) fluctuations.  
At high enough temperatures these fluctuations can be viewed as a quenched Gaussian
random field with finite correlation length. This generalized DMFT+$\Sigma_{\bf k}$ approach
is used for the numerical solution of 
the weakly doped one--band Hubbard model with repulsive Coulomb interaction on a
square lattice with nearest and next nearest neighbour hopping.  
The effective single impurity problem in this generalized DMFT+$\Sigma_{\bf k}$
is solved by numerical renormalization 
group (NRG).  Both types of strongly correlated metals, namely (i) doped Mott 
insulator and (ii) the case of bandwidth $W\lesssim U$ ($U$ --- value of local 
Coulomb interaction) are considered. Densities of states, spectral functions
and ARPES spectra calculated within DMFT+$\Sigma_{\bf k}$
show a pseudogap formation near the Fermi level of the quasiparticle band.  
\end{abstract}


\maketitle

\newpage

\section{Introduction}

Among the numerous anomalies of the normal phase of high--temperature
superconductors the observation of a pseudogap
in the electronic spectrum of underdoped copper oxides \cite{Tim,MS}
is especially interesting.  
Despite continuing discussions on the nature of
the pseudogap, the preferable ``scenario'' for its formation 
is most likely based on the model of strong scattering of the charge
carriers by short--ranged antiferromagnetic (AFM, SDW) spin fluctuations
\cite{MS,Pines}. In momentum representation this scattering transfers 
momenta of the order of ${\bf Q}=(\frac{\pi}{a},\frac{\pi}{a})$ 
($a$ --- lattice constant of two dimensional lattice). 
This leads to the formation of structures in the one-particle spectrum, 
which are precursors of the changes in the spectra due
to long--range AFM order (period doubling).
As a result we obtain non--Fermi liquid like behavior (dielectrization)
of the spectral density in the vicinity of the so called ``hot spots'' on the
Fermi surface, appearing at intersections of the Fermi surface 
with antiferromagnetic Brillouin zone boundary \cite{MS}.

Within this spin--fluctuation scenario a simplified model of the pseudogap 
state was studied \cite{MS,Sch,KS} under the assumption that the scattering
by dynamic
spin fluctuations can be reduced for high enough temperatures
to a static Gaussian random field (quenched disorder) of pseudogap fluctuations.
These fluctuations are defined by a characteristic scattering vector from the 
vicinity of ${\bf Q}$,  with a width 
determined by the inverse correlation length of short--range
order $\kappa=\xi^{-1}$. 

Undoped cuprates are antiferromagnetic Mott insulators with
$U\gg W$ ($U$ --- value of local Coulomb interaction, $W$ --- bandwidth of
non--interacting band), so that
correlation effects are very important.
It is thus clear that the electronic properties of underdoped 
(and probably also optimally doped) cuprates are governed by strong electronic 
correlations too, so that these systems are typical strongly correlated
metals. Two types of correlated metals can be distinguished:
(i) the doped Mott insulator and (ii) the bandwidth controlled 
correlated metal $W\approx U$. Both types will be considered in this paper.

A state of the art tool to describe such correlated  systems
is the dynamical mean--field theory (DMFT)
\cite{MetzVoll89,vollha93,pruschke,georges96,PT}.
The characteristic features of correlated systems within the DMFT
are the formation of incoherent structures, the so-called Hubbard bands,
split by the Coulomb interaction $U$, and a
quasiparticle (conduction) band near the Fermi level dynamically 
generated by the local correlations 
\cite{MetzVoll89,vollha93,pruschke,georges96,PT}.

Unfortunately, the DMFT is not useful to the study the
``antiferromagnetic'' scenario of pseudogap formation in strongly
correlated metals. This is due to the basic approximation of the DMFT, which
amounts to the complete neglect of non-local dynamical correlation effects.
The aim of the present paper is to formulate an approach overcoming this 
difficulty.

The paper is organized as follows:
In section \ref{leng_intro} we present a derivation of the
self--consistent generalization we call DMFT+$\Sigma_{\bf k}$ 
which includes short-ranged dynamical correlations to some extent.
Section \ref{kself} describes the construction of the
{\bf k}--dependent self--energy, and some 
computational details are presented in section \ref{compdet}.  
Results and a discussion are given in the sections \ref{results} and 
\ref{concl}.

\section{Introducing length scale into DMFT: DMFT+$\Sigma_{\bf k}$ approach}
\label{leng_intro}

Basic shortcoming of traditional DMFT approach
\cite{MetzVoll89,vollha93,pruschke,georges96,PT}
is the neglect of momentum dependence of electron self--energy.
This approximation in principle allows for an exact solution of correlated
electron systems fully preserving the local part of the dynamics introduced
by electronic correlations.
To include non--local effects, while remaining within the usual ``impurity
analogy'', we propose the following procedure. To be definite, let us consider
a standard one-band Hubbard model from now on. The extension to multi-orbital or
multi-band models is straightforward.
The major assumption of our approach is that the lattice
and Matsubara ``time'' Fourier transform of the single-particle Green function 
can be written as:
\begin{equation}
G_{\bf k}(i\omega)=\frac{1}{i\omega+\mu-\varepsilon({\bf k})-\Sigma(i\omega)
-\Sigma_{\bf k}(i\omega)},\qquad \omega=\pi T(2n+1),
\label{Gk}
\end{equation}
where $\Sigma(i\omega)$ is the {\em local} contribution to self--energy,
surviving in the DMFT, while $\Sigma_{\bf k}(i\omega)$
is some momentum dependent part. We suppose that
this last contribution is due to either electron interactions with some
``additional'' collective modes or order parameter fluctuations, or may be
due to non--local contributions within the Hubbard model itself. The assumed
additive form of self--energy $\Sigma(i\omega)+\Sigma_{\bf k}(i\omega)$
corresponds to neglect of possible interference
of these local and non--local contributions.

The self--consistency equations of our generalized DMFT+$\Sigma_{\bf k}$ approach
are formulated as follows:
\begin{enumerate}
\item{Start with some initial guess of {\em local} self--energy
$\Sigma(i\omega)$, e.g. $\Sigma(i\omega)=0$}.  
\item{Construct $\Sigma_{\bf k}(i\omega)$ within some (approximate) scheme, 
taking into account interactions with collective modes or order parameter
fluctuations which in general can depend on $\Sigma(i\omega)$
and $\mu$.} 
\item{Calculate the local Green function  
\begin{equation}
G_{ii}(i\omega)=\frac{1}{N}\sum_{\bf k}\frac{1}{i\omega+\mu
-\varepsilon({\bf k})-\Sigma(i\omega)-\Sigma_{\bf k}(i\omega)}.
\label{Gloc}
\end{equation}
}
\item{Define the ``Weiss field''
\begin{equation}
{\cal G}^{-1}_0(i\omega)=\Sigma(i\omega)+G^{-1}_{ii}(i\omega).
\label{Wss}
\end{equation}
}
\item{Using some ``impurity solver'' to calculate the single-particle Green function for the 
effective Anderson impurity problem, defined by 
integral 
\begin{equation}
G_{d}(\tau-\tau')=\frac{1}{Z_{\text{eff}}}
\int Dc^+_{i\sigma}Dc_{i\sigma}
c_{i\sigma}(\tau)c^+_{i\sigma}(\tau')\exp(-S_{\text{eff}})
\label{AndImp}
\end{equation}
with effective action for a fixed site (``impurity'') $i$
\begin{equation}
S_{\text{eff}}=-\int_{0}^{\beta}d\tau_1\int_{0}^{\beta}
d\tau_2c_{i\sigma}(\tau_1){\cal G}^{-1}_0(\tau_1-\tau_2)c^+_{i\sigma}(\tau_2)
+\int_{0}^{\beta}d\tau Un_{i\uparrow}(\tau)n_{i\downarrow}(\tau)\;\;,
\label{Seff}
\end{equation}
$Z_{\text{eff}}=\int Dc^+_{i\sigma}Dc_{i\sigma}\exp(-S_{\text{eff}})$, and
$\beta=T^{-1}$. This step produces a {\em new} set of values $G^{-1}_{d}(i\omega)$.}
\item{Define a {\em new} local self--energy
\begin{equation}
\Sigma(i\omega)={\cal G}^{-1}_0(i\omega)-
G^{-1}_{d}(i\omega).
\label{StS}
\end{equation}
}
\item{Using this self--energy as ``initial'' one in step 1, continue the procedure
until (and if) convergency is reached to obtain
\begin{equation}
G_{ii}(i\omega)=G_{d}(i\omega).
\label{G00}
\end{equation}
}
\end{enumerate}
Eventually, we get the desired Green function in the form of (\ref{Gk}),
where $\Sigma(i\omega)$ and $\Sigma_{\bf k}(i\omega)$ are those appearing
at the end of our iteration procedure.
A more detailed derivation of this scheme within a diagrammatic approach 
is  given in the Appendix \ref{A}.

\section{Construction of {\bf k}--dependent self--energy}
\label{kself}
For the momentum dependent part of the single-particle self--energy we concentrate
on the effects of scattering of electrons from collective short-range
SDW--like antiferromagnetic spin (or CDW--like charge) fluctuations.
To calculate $\Sigma_{\bf k}(i\omega)$ for an electron moving in the quenched
random field of (static) Gaussian spin (or charge) fluctuations with dominant
scattering momentum transfers from the vicinity of some characteristic
vector ${\bf Q}$, we use a slightly generalized version of the recursion procedure 
proposed in Refs.~\cite{MS79,Sch,KS} and take into account {\em all} Feynman 
diagrams describing the scattering of electrons by this random field. 
Then the desired self--energy is given by
\begin{equation}
\Sigma_{\bf k}(i\omega)=\Sigma_{n=1}(i\omega{\bf k})
\label{Sk}
\end{equation}
with
\begin{equation}
\Sigma_{n}(i\omega{\bf k})=\Delta^2\frac{s(n)}
{i\omega+\mu-\Sigma(i\omega)
-\varepsilon_n({\bf k})+inv_n\kappa-\Sigma_{n+1}(i\omega{\bf k})}\;\;. 
\label{rec}
\end{equation} 
The quantity $\Delta$ characterizes the energy scale and
$\kappa=\xi^{-1}$ is the inverse correlation length of short range
SDW (CDW) fluctuations, $\varepsilon_n({\bf k})=\varepsilon({\bf k+Q})$ and 
$v_n=|v_{\bf k+Q}^{x}|+|v_{\bf k+Q}^{y}|$ 
for odd $n$ while $\varepsilon_n({\bf k})=\varepsilon({\bf k})$ and $v_{n}=
|v_{\bf k}^x|+|v_{\bf k}^{y}|$ for even $n$. The velocity projections
$v_{\bf k}^{x}$ and $v_{\bf k}^{y}$ are determined by usual momentum derivatives
of the ``bare'' electronic energy dispersion $\varepsilon({\bf k})$. Finally,
$s(n)$ represents a combinatorial factor with
\begin{equation}
s(n)=n
\label{vcomm}
\end{equation}
for the case of commensurate charge (CDW type) fluctuations with
${\bf Q}=(\pi/a,\pi/a)$ \cite{MS79}. 
For incommensurate CDW fluctuations \cite{MS79} one finds
\begin{equation} 
s(n)=\left\{\begin{array}{cc}
\frac{n+1}{2} & \mbox{for odd $n$} \\
\frac{n}{2} & \mbox{for even $n$}.
\end{array} \right.
\label{vinc}
\end{equation}
If we want to take into account the (Heisenberg) spin structure of interaction with 
spin fluctuations
in  ``nearly antiferromagnetic Fermi--liquid'' 
(spin--fermion (SF) model Ref.~\cite{Sch}),
the combinatorics of diagrams becomes more complicated.
Spin--conserving scattering processes obey commensurate combinatorics,
while spin--flip scattering is described by diagrams of incommensurate
type (``charged'' random field in terms of Ref.~\cite{Sch}). In this model
the recursion relation for the single-particle Green function is again given by
(\ref{rec}), 
but the combinatorial factor $s(n)$ now acquires the following form \cite{Sch}:
\begin{equation} 
s(n)=\left\{\begin{array}{cc}
\frac{n+2}{3} & \mbox{for odd $n$} \\
\frac{n}{3} & \mbox{for even $n$}.
\end{array} \right.
\label{vspin}
\end{equation}
Obviously, with this procedure we introduce an important length scale $\xi$ 
not present in standard DMFT. Physically this scale mimics the effect of short-range
(SDW or CDW) correlations within fermionic ``bath'' surrounding the effective Anderson
of the DMFT impurity. We expect that such a length-scale will lead to a competition
between local and non-local physics.

An important aspect of the theory is that both parameters $\Delta$ and $\xi$ can in
principle be calculated from the microscopic model at hand. For example, using the
two--particle selfconsistent approach of Ref.~\cite{VT} with the approximations
introduced in Refs.~\cite{Sch,KS}, one can derive within the
standard Hubbard model the following microscopic expression for $\Delta$:
\begin{eqnarray} 
\Delta^2=\frac{1}{4}U^2\frac{<n_{i\uparrow}n_{i\downarrow}>}
{<n_{i\uparrow}><n_{i\downarrow}>}[<n_{i\uparrow}>+<n_{i\downarrow}>
-2<n_{i\uparrow}n_{i\downarrow}>]=\nonumber\\
=U^2\frac{<n_{i\uparrow}n_{i\downarrow}>}{n^2}<(n_{i\uparrow}
-n_{i\downarrow})^2>
=\nonumber\\
=U^2\frac{<n_{i\uparrow}n_{i\downarrow}>}{n^2}\frac{1}{3}<{\vec S}_i^2>,
\label{DeltHubb}
\end{eqnarray}
where we consider only scattering from antiferromagnetic spin fluctuations.
The different local quantities -- spin fluctuation $<{\vec S}_i^2>$,  density
$n$ and  double occupancy
$<n_{i\uparrow}n_{i\downarrow}>$ -- can easily be calculated within the 
standard DMFT \cite{georges96}.
A detailed derivation of (\ref{DeltHubb}) and computational
results for $\Delta$ obtained by DMFT using quantum Monte--Carlo (QMC)
to solve the effective impurity problem are presented in Appendix \ref{B}.
A corresponding microscopic expressions for the correlation length $\xi=\kappa^{-1}$
can also be derived within the two--particle self--consistent
approach \cite{VT}. However, we expect those results for $\xi$ to be less reliable,
because this approach is valid only for relatively small (or medium) 
values of $U/t$.
Thus, in the following we will consider both $\Delta$ and especially $\xi$
as some phenomenological parameters to be determined from experiments.

\section{Results and discussion}
\label{results}

\subsection{Computation details}
\label{compdet}
In the following, we want to discuss results for a standard one-band
Hubbard model on a square lattice. With nearest ($t$) and
next nearest ($t'$) neighbour hopping integrals the dispersion
then reads
\begin{equation}
\varepsilon({\bf k})=-2t(\cos k_xa+\cos k_ya)-4t'\cos k_xa\cos k_ya\;\;,
\label{spectr}
\end{equation}
where $a$ is the lattice constant.
The correlations are introduced by a repulsive local two-particle interaction $U$.
%
We choose as energy scale the nearest neighbour hopping integral $t$
and as length scale the lattice constant $a$.

For a square lattice the bare bandwidth is $W=8t$.
To study a strongly correlated metallic state obtained as doped Mott insulator
we use $U=40t$ as value for the Coulomb interaction and a filling $n=0.8$ (hole 
doping).  The correlated metal in the case of $W\gtrsim U$ 
is realized via $U=4t$ and two fillings: half--filling ($n=1.0$) and $n=0.8$ (hole 
doping). As typical values for $\Delta$ we choose $\Delta=t$ and $\Delta=2t$
(actually as approximate limiting values --- cf. Appendix~\ref{B})
and for the correlation length $\xi=2a$ and $\xi=10a$ (motivated mainly by 
experimental data for cuprates~\cite{MS,Sch}).

The DMFT maps the lattice problem onto an effective, 
self--consistent impurity defined by Eqs. (\ref{AndImp})-(\ref{Seff}).  
In our work we employ as ``impurity solvers'' two 
reliable numerically exact methods --- quantum Monte--Carlo (QMC)~\cite{QMC} 
and numerical renormalization group (NRG) \cite{NRG,BPH}.
Calculations were done for the case $t'=0$ and 
$t'/t$=-0.4 (more or less typical for cuprates)
at two different temperatures $T=0.088t$ and $T=0.356t$ (for NRG 
computations)~\footnote{
Discretization parameter $\Lambda$=2, number of
low energy states after truncation 1000, cut-off near Fermi energy 10$^{-6}$,
broadening parameter b=0.6.}
.
QMC computations of double occupancies as functions of 
filling were done at temperatures $T=0.1t$ and $T=0.4t$ 
~\footnote{
Number warm-up sweeps 30000, number of QMC sweeps 200000,
number of imaginary time slices 40.}. 

\subsection{Generalized DMFT+$\Sigma_{\bf k}$ approach: densities of states}

\begin{figure}[htb]
\includegraphics[clip=true,width=0.6\textwidth]{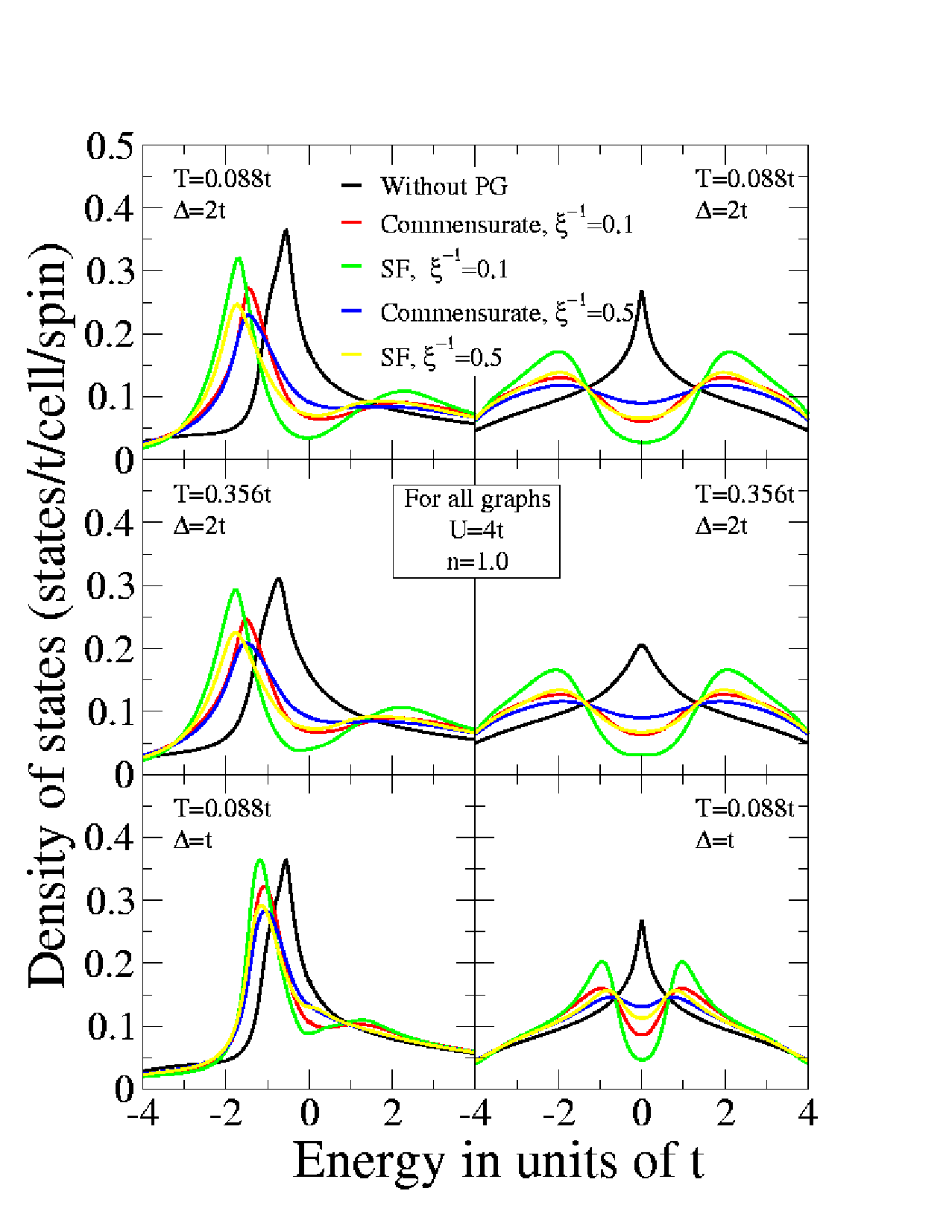}
\caption{Comparison of DOSs obtained from
DMFT(NRG)+$\Sigma_{\bf k}$ calculations for different
combinatorical factors (SF --- spin--fermion model, commensurate),
inverse correlation lengths
($\xi^{-1}$) in units of the lattice constant, temperatures ($T$) and 
values of the pseudogap potential ($\Delta$).
Left column corresponds to $t'/t=-0.4$, right column to $t'=0$.
In all graphs the Coulomb interaction is $U=4t$ and $n=1$.
The Fermi level corresponds to zero.}
\label{DOS_4t_n1}
\end{figure}

\begin{figure}[htb]
\includegraphics[clip=true,width=0.6\textwidth]{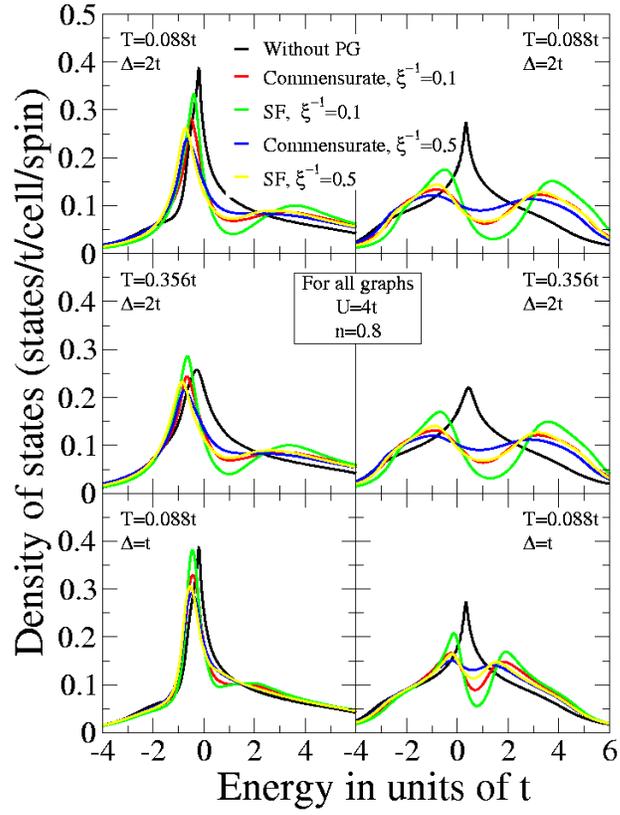}
\caption{Comparison of DOSs obtained from
DMFT(NRG)+$\Sigma_{\bf k}$ calculations for a filling $n=0.8$, other parameters
as in Fig.~\ref{DOS_4t_n1}.}
\label{DOS_4t_n08}
\end{figure}

Let us start the discussion of our results
obtained within our generalized DMFT+$\Sigma_{\bf k}$ approach
with the densities of states (DOSs) for the case of small (relative
to bandwidth) Coulomb interaction $U=4t$ with and without pseudogap fluctuations.
As already discussed in the introduction, the characteristic feature of the strongly
correlated metallic state
is the coexistence of lower and upper Hubbard bands split by the value of $U$
with a quasiparticle peak at the Fermi level.
Since at half--filling the bare DOS of the square lattice has a Van--Hove singularity 
at the Fermi level ($t'=0$) or close to it (in case of $t'/t=-0.4$) 
one cannot treat a peak on the Fermi level simply as a quasiparticle peak. In fact,
there are two contributions to this peak from (i) 
the quasiparticle peak appearing in strongly correlated metals due to many-body effects
and (ii) the smoothed Van--Hove singularity from the bare DOS~\cite{VHS}.  
In Figs.~\ref{DOS_4t_n1} and~\ref{DOS_4t_n08} we show the corresponding DMFT(NRG) DOSs 
without pseudogap fluctuations as black lines for $n=1$ and $n=0.8$ for both
bare dispersions $t'/t=-0.4$ (left panels) and for $t'=0$ (right panels) for two
different  temperatures $T=0.356t$ (middle panels) and $T=0.088t$ (upper and lower
panels).
The remaining curves in Figs.~\ref{DOS_4t_n1} and~\ref{DOS_4t_n08} represent results
for the DOSs with non-local fluctuations switched on.
The fluctuation amplitudes are taken as $\Delta=2t$ (upper and and middle panels) 
and $\Delta=t$ (low panels). For all sets of parameters one can see that the
introduction of non-local fluctuations into the calculation leads to the formation of 
pseudogap in the quasiparticle peak.

The behaviour of the pseudogaps in the DOSs has some common features.
For example, for $t'$=0 at half--filling (Fig. 
\ref{DOS_4t_n1}, right column) we find that the pseudogap is most pronounced.
For $n=0.8$ (Fig. \ref{DOS_4t_n08}, right column) the picture is almost the same
but slightly asymmetric. The width of the pseudogap (the distance between peaks closest
to Fermi level) appears to be of the order of $\sim 2\Delta$ here. 
Decreasing the value of $\Delta$ from $2t$ to $t$ leads to a pseugogap that is
correspondingly twice smaller and in addition more shallow. When one uses the
combinatorial factors corresponding to the
spin--fermion model (Eq.(\ref{vspin})), we find that the pseudogap becomes more 
pronounced than in the case of commensurate charge fluctuations (combinatorial factors
of Eq. (\ref{vinc})). 
The influence of the correlation length $\xi$ can be seen is also as expected. Changing
$\xi^{-1}$ from $\xi^{-1}=0.1$ to $\xi^{-1}=0.5$, i.e.\ decreasing the range
of the non-local fluctuations, slightly washes out the pseudogap.  
Also, increasing the temperature from $T=0.088t$ to $T=0.356t$ leads to a general
broadening of the structures in the DOSs.  These observations remain at least
qualitatively
valid for $t'/t=-0.4$ (Figs. \ref{DOS_4t_n1} and \ref{DOS_4t_n08}, left columns)
with an additional asymmetry due to the next-nearest neighbour hopping.
Noteworthy is however the fact that for $t'/t=-0.4$ and $\xi ^{-1}=0.5$ the 
pseudogap has almost disappeared for the temperatures studied here.  An also very
remarkable point is the similarity of the results 
obtained with the generalized DMFT+$\Sigma_{\bf k}$ approach with $U=4t$ (smaller
than the  bandwidth $W$) to those obtained earlier without Hubbard--like Coulomb 
interactions \cite{Sch,KS}.

\begin{figure}[htb]
\includegraphics[clip=true,width=0.6\textwidth,angle=270]{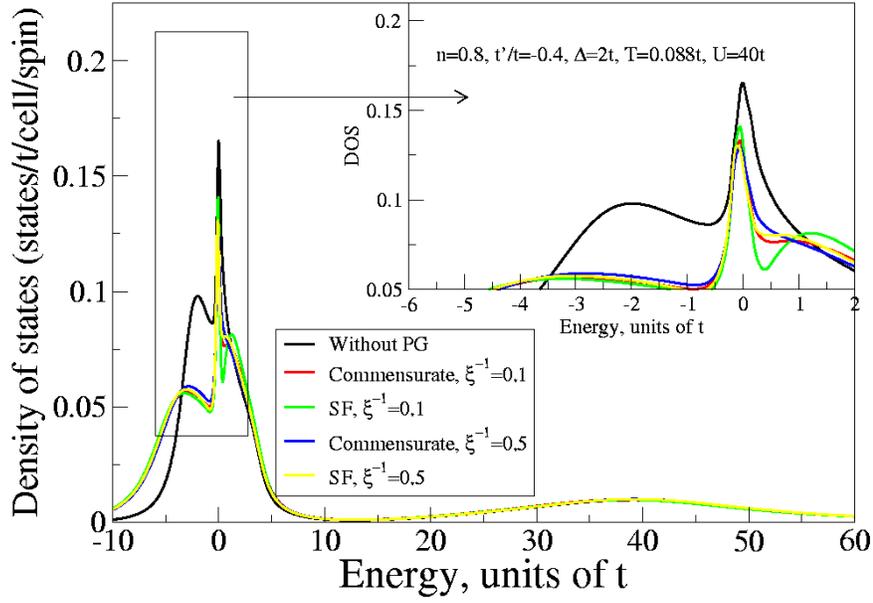}
\includegraphics[clip=true,width=0.6\textwidth,angle=270]{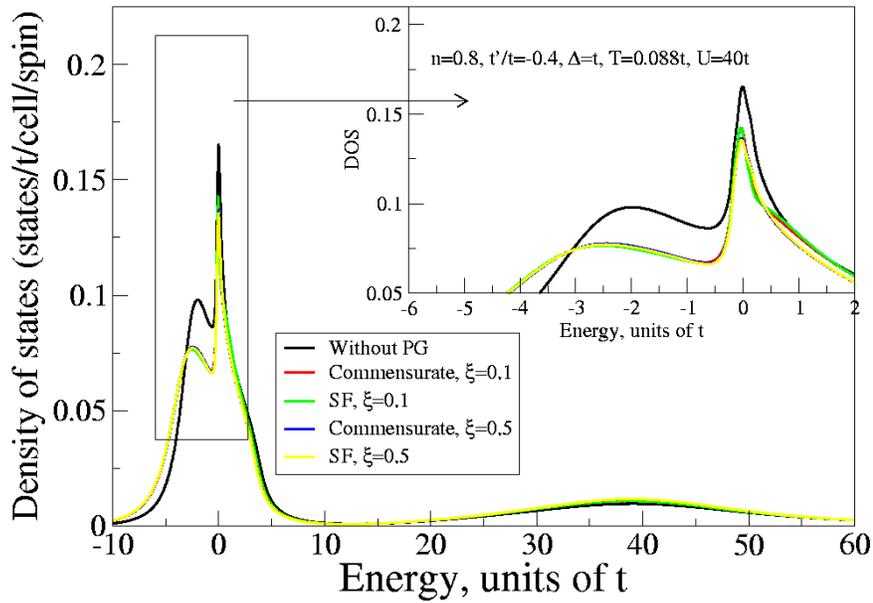}
\caption{Comparison of DOSs obtained from
DMFT(NRG)+$\Sigma_{\bf k}$ calculations for $t'/t=-0.4$, $T=0.088t$, $U=40t$ and filling $n=0.8$.}
\label{dos_40t_04}
\end{figure}

\begin{figure}[htb]
\includegraphics[clip=true,width=0.6\textwidth,angle=270]{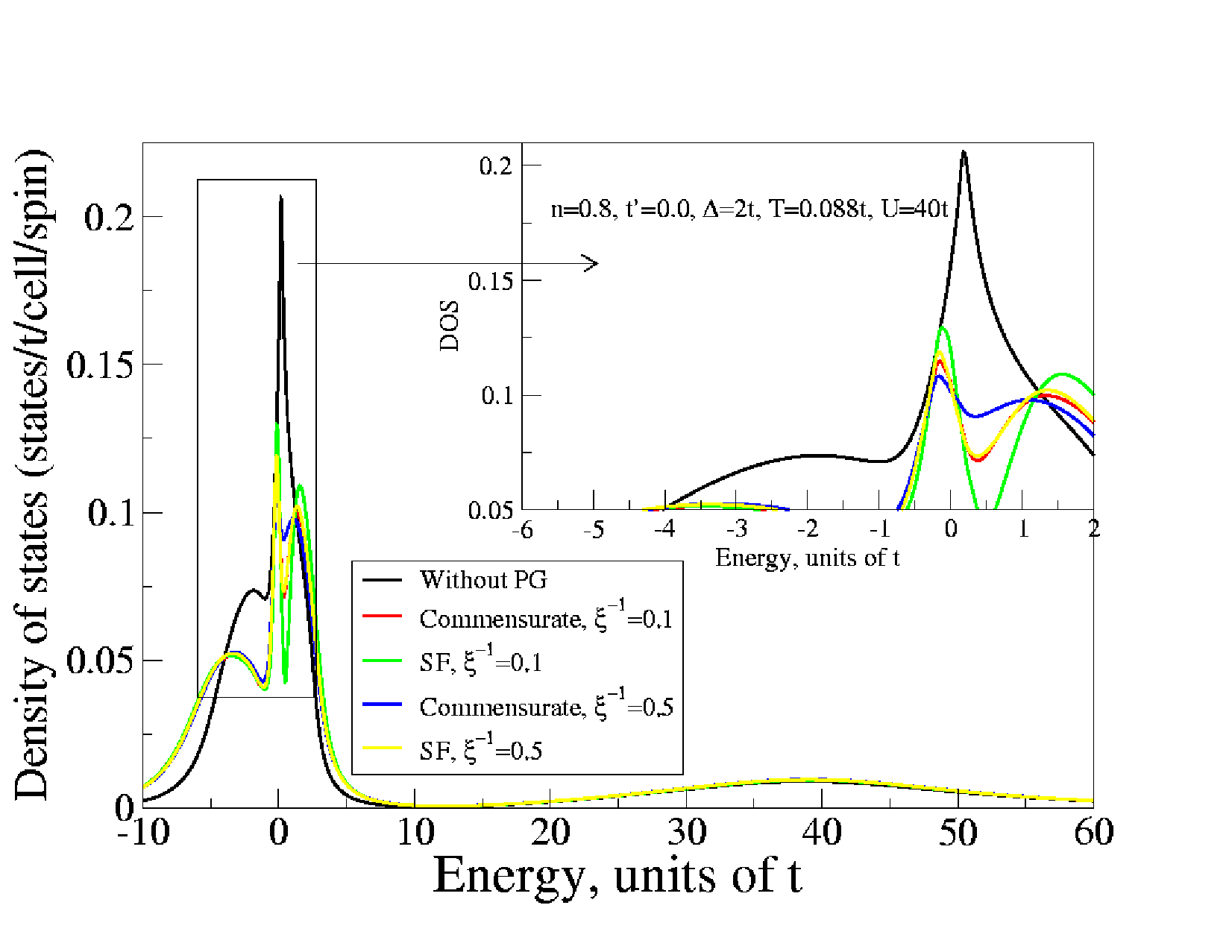}
\includegraphics[clip=true,width=0.6\textwidth,angle=270]{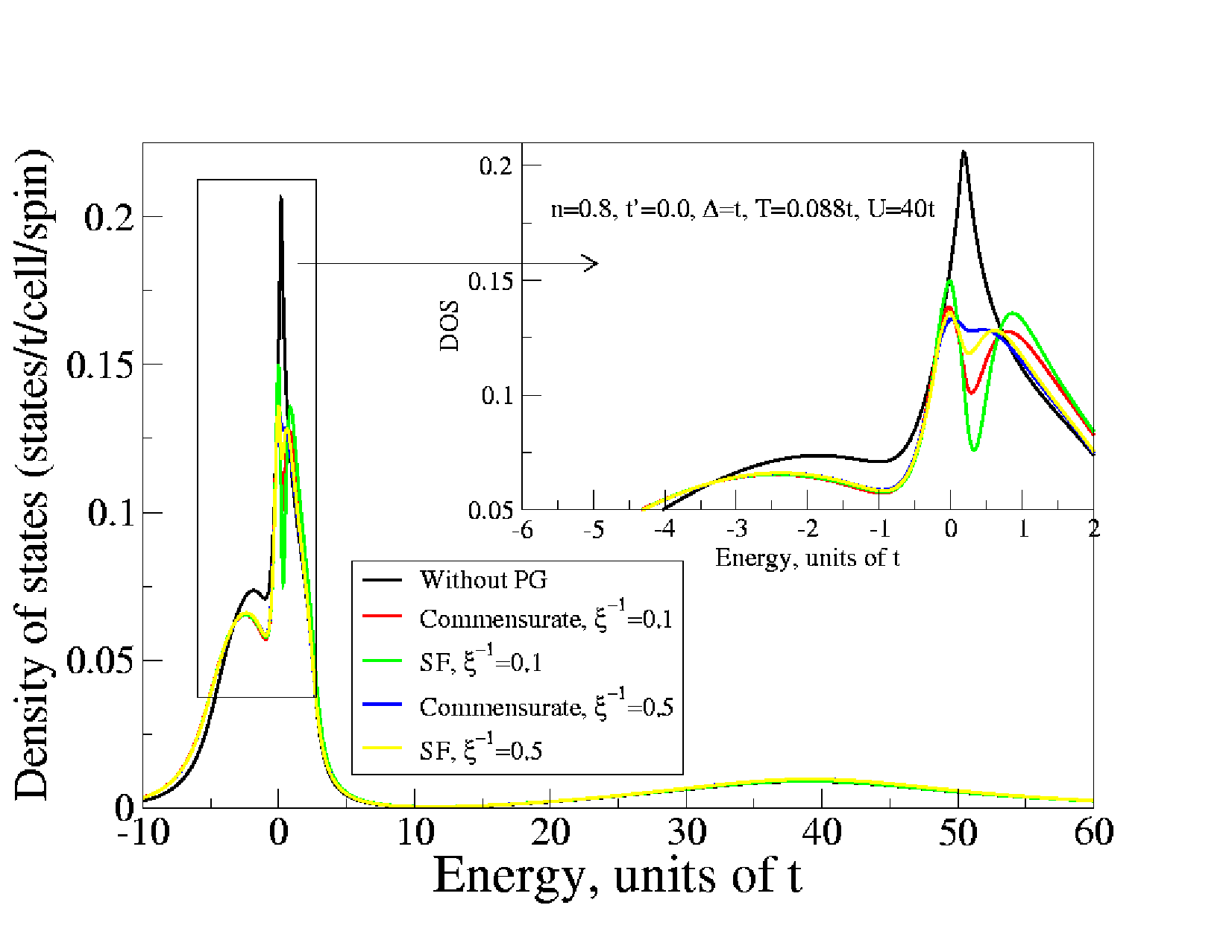}
\caption{Comparison of DOSs obtained from
DMFT(NRG)+$\Sigma_{\bf k}$ calculations for $t'=0$, other parameters as in Fig.~\ref{dos_40t_04}.}
\label{dos_40t_0}
\end{figure}

Let us now consider the case of a doped Mott insulator.  The model parameters
are again $t'=0$ and $t'/t=-0.4$ with filling 
$n=0.8$, but the Coulomb interaction strength is set to $U=40t$ for our DMFT+$\Sigma_{\bf k}$ 
calculations.  Characteristic features of the DOS for such a strongly correlated metal 
are a strong separation of lower and upper Hubbard bands and a Fermi level 
crossing by the lower Hubbard band (for non--half--filled case). 
Without non-local fluctuations the quasi-particle peak is again formed at the Fermi
level; but now the upper Hubbard band is far to the 
right and does not touch the quasiparticle peak (as it was for the case of small 
Coulomb interactions). DOSs without non-local fluctuations are again presented as 
black lines on Fig.~\ref{dos_40t_0} ($t'=0$) and Fig.~\ref{dos_40t_04} 
($t'/t=-0.4$).

With rather strong non-local fluctuations $\Delta =2t$,
a pseudogap appears in the middle of quasiparticle peak. In addition we observe that
the lower Hubbard band is slightly broadened by fluctuation effects.
Qualitative behaviour of the pseudogap anomalies is again similar to those
described above for the case of $U=4t$, e.g.\ a decrease of $\xi$ makes the pseudogap
less pronounced, reducing $\Delta$ from $\Delta =2t$ to $\Delta =t$ narrows of the
pseudogap and
also makes it more shallow etc.. Note that for the doped Mott--insulator we find that
the pseudogap is remarkably more pronounced for the SDW--like fluctuations
than for CDW--like fluctuations.

There are, however, quite clear differences to the case with $U=4t$. For example,
the width of the pseudogap appears to be much smaller than $2\Delta$, which we
attribute to the fact that the quasiparticle peak itself is actually rather narrow
now. 

\subsection{Generalized DMFT+$\Sigma_{\bf k}$ approach:
spectral functions $A(\omega,{\bf k})$}

\begin{figure}[htb]
\includegraphics[clip=true,width=0.7\textwidth,angle=270]{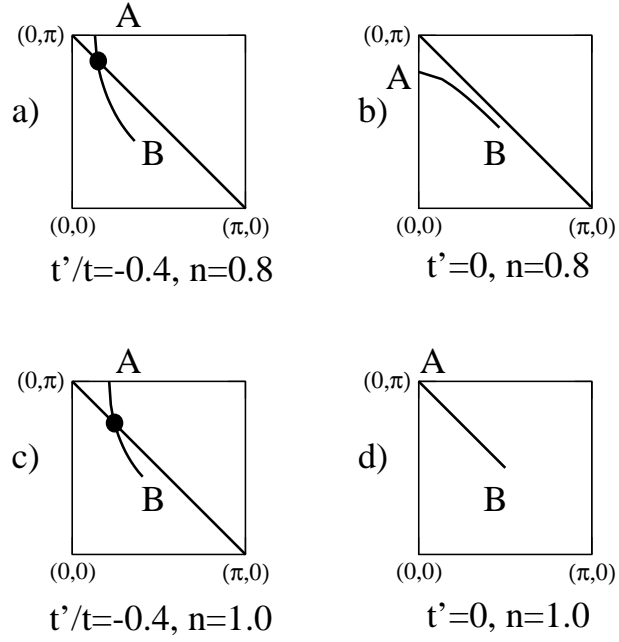}
\caption{1/8-th of the bare Fermi surfaces for
the different occupancies $n$ and combinations ($t,t'$)
used for the calculation of spectral functions $A({\bf k},\omega)$.
The diagonal line corresponds to the antiferromagnetic Brillouin zone
boundary at half--filling for a
square lattice with nearest-neighbours hopping only. The full circle marks 
the so-called ``hot--spot''.}
\label{FS_shapes}
\end{figure}

In the previous subsections we discussed the densities of states obtained
self--consistently by the DMFT+$\Sigma_{\bf k}$ approach. Once we get a
self--consistent solution of the DMFT+$\Sigma_{\bf k}$ equations with
non-local fluctuations we can of course also compute the spectral functions
$A(\omega,{\bf k})$
\begin{equation}
A(\omega,{\bf k})=-\frac{1}{\pi}{\rm Im}\frac{1}{\omega+\mu
-\varepsilon({\bf k})-\Sigma(\omega)-\Sigma_{\bf k}(\omega)},
\label{specf}
\end{equation}
where self--energy $\Sigma(\omega)$ and chemical potential $\mu$
are calculated self--consistently
as described in Sec. \ref{leng_intro}. 
To plot $A(\omega,{\bf k})$ we choose ${\bf k}$--points along the
``bare'' Fermi surfaces for different types of
lattice spectra and fillings. In Fig. \ref{FS_shapes}
one can see corresponding shapes of these ``bare'' Fermi surfaces (presented 
are only 1/8-th parts of the Fermi surfaces within the first quadrant of the
Brillouin zone).

In the following we concentrate mainly on the case $U=4t$ and
filling $n=0.8$ (Fermi surface of Fig. \ref{FS_shapes}(a)). The
corresponding spectral functions $A(\omega,{\bf k})$ are depicted in
Fig.~\ref{sf_U4t_n08}. 
When $t'/t=-0.4$ (upper row), the spectral function close to the Brillouin zone diagonal
(point B) has the typical Fermi--liquid behaviour, consisting of a rather sharp peak
close to the Fermi level.
In the case of SDW--like fluctuations 
this peak is shifted down in energy by about $-0.5t$ (left upper corner).
In the vicinity of the ``hot--spot'' the shape of $A(\omega,{\bf k})$ is 
completely modified. Now $A(\omega,{\bf k})$ becomes double--peaked and
non--Fermi--liquid--like.  Directly at the ``hot--spot'', $A(\omega,{\bf k})$ 
for SDW--like  fluctuations has two equally intensive peaks situated 
symmetrically around the Fermi level and split from each other by $\sim 
1.5\Delta$ Refs.~\cite{Sch,KS}.  For commensurate CDW--like fluctuations the
spectral function in the 
``hot--spot'' region has one broad peak centred at the Fermi level 
with width $\sim \Delta$. Such a merging of the two peaks at the ``hot--spot'' 
for commensurate fluctuations was previously observed in Ref. \cite{KS}.
However close to point A  this type of fluctuations also produces a
double--peak structure in the spectral function.

\begin{figure}[htb]
\includegraphics[clip=true,width=0.6\textwidth]{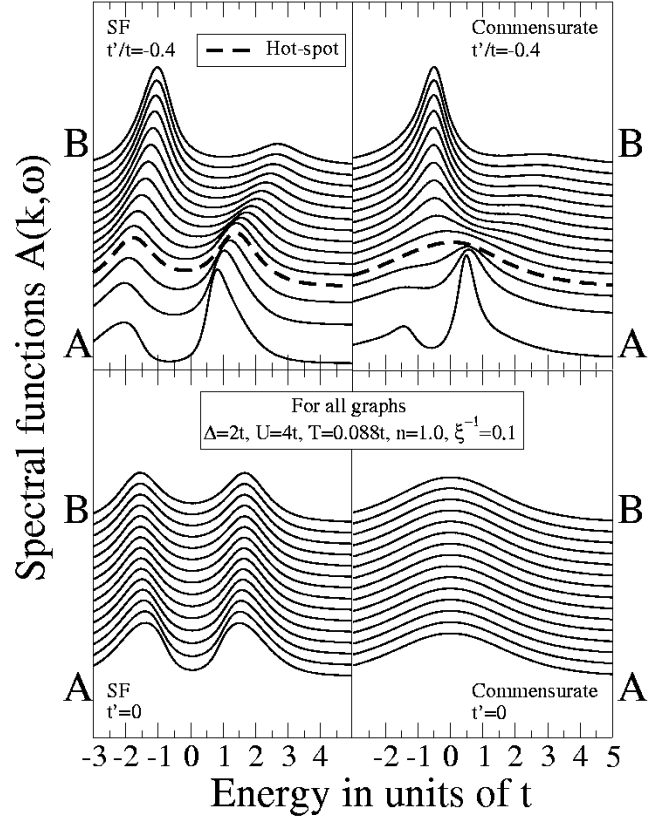}
\caption{Spectral functions $A({\bf k},\omega)$
obtained from DMFT(NRG)+$\Sigma_{\bf k}$ calculations along the directions
shown in Fig.~\protect\ref{FS_shapes}.
Model parameters were chosen as 
$U=4t$, $n=1.0$, $\Delta=2t$, $\xi^{-1}=0.1$ 
and temperature $T=0.088t$. 
The ``hot--spot'' {\bf k}-point is marked as fat dashed line.
The Fermi level corresponds to zero.}
\label{sf_U4t_n1}
\end{figure}

\begin{figure}[htb]
\includegraphics[clip=true,width=0.6\textwidth]{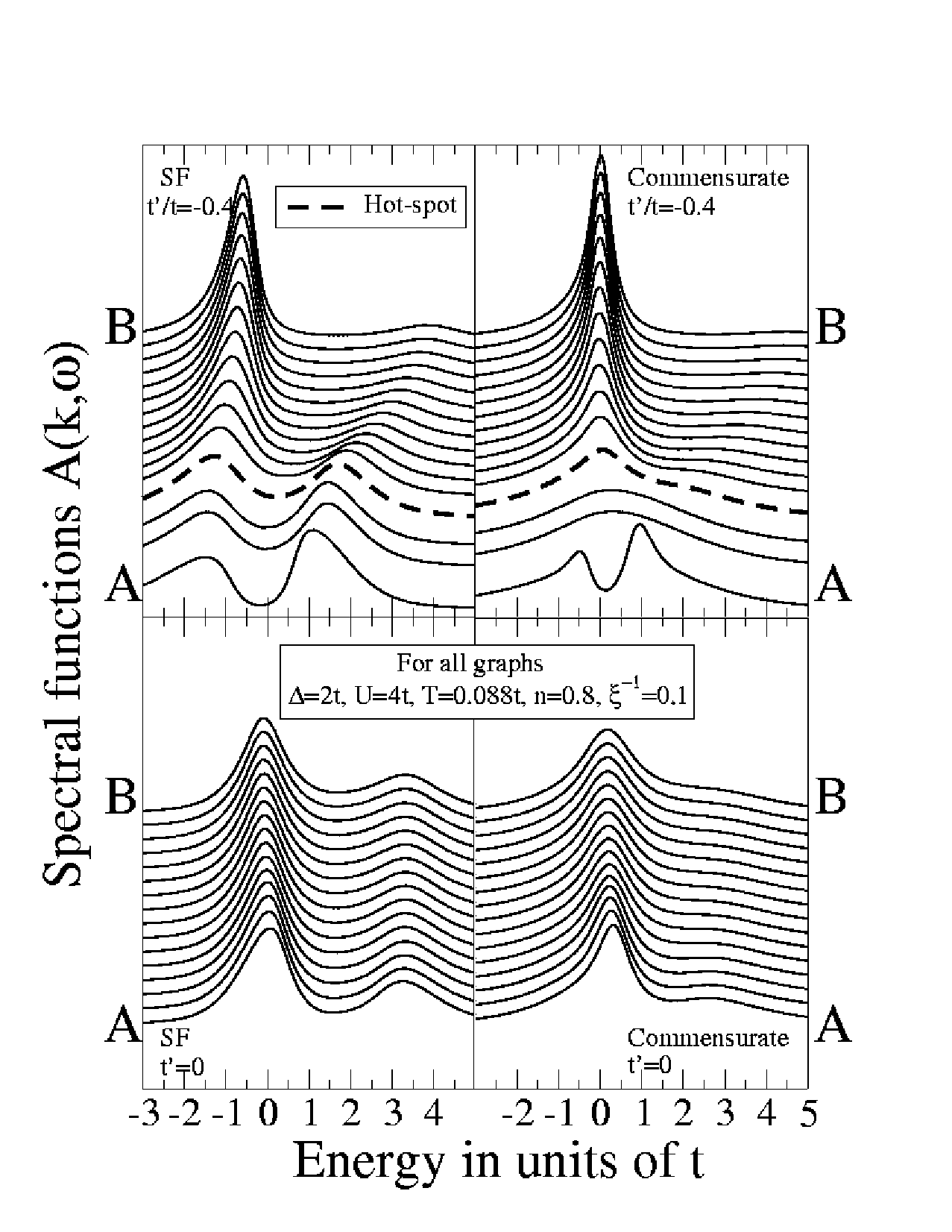}
\caption{Spectral functions $A({\bf k},\omega)$ 
obtained from the DMFT(NRG)+$\Sigma_{\bf k}$ calculations for $n=0.8$,
other parameters as in Fig.~\ref{sf_U4t_n1}.}
\label{sf_U4t_n08}
\end{figure}

In Fig. \ref{sf_U4t_n1} we show spectral functions for the case of
$U=4t$ at half--filling ($n=1$) (Fermi surface of Fig. \ref{FS_shapes} (c), 
(d)).  For $t'/t=-0.4$ (upper row of Fig. \ref{sf_U4t_n1}) these are
similar to those just discussed for $n=0.8$. However, the
pseugogap is more pronounced here and remains open even close 
to the antiferromagnetic Fermi surface boundary (point B) for SDW fluctuations.

In the lower panel of Fig.~\ref{sf_U4t_n08} we show spectral functions
for 20\% hole doping ($n=0.8$) and the case of $t'=0$ (Fermi surface from 
Fig. \ref{FS_shapes}(b)). 
Since the Fermi surface now is close to the
antiferromagnetic zone boundary, the pseudogap anomalies are rather strong and 
almost non--dispersive along the Fermi surface.  At half--filling (Fig. 
\ref{sf_U4t_n08}, lower panel) and for $t'=0$ the Fermi surface
(Fig.~\ref{FS_shapes}(d)) actually coincides with antiferromagnetic zone
boundary.  In this case the the whole Fermi surface is 
the ``hot--region'' (perfect ``nesting'').  
The spectral functions are now symmetric around the Fermi level.  
For SDW--like fluctuations there are two peaks 
split by $\sim 1.5\Delta$. Again, CDW--like fluctuations give just one 
peak centred at the Fermi level with width $\sim \Delta$.

\begin{figure}[htb]
\includegraphics[clip=true,width=0.6\textwidth]{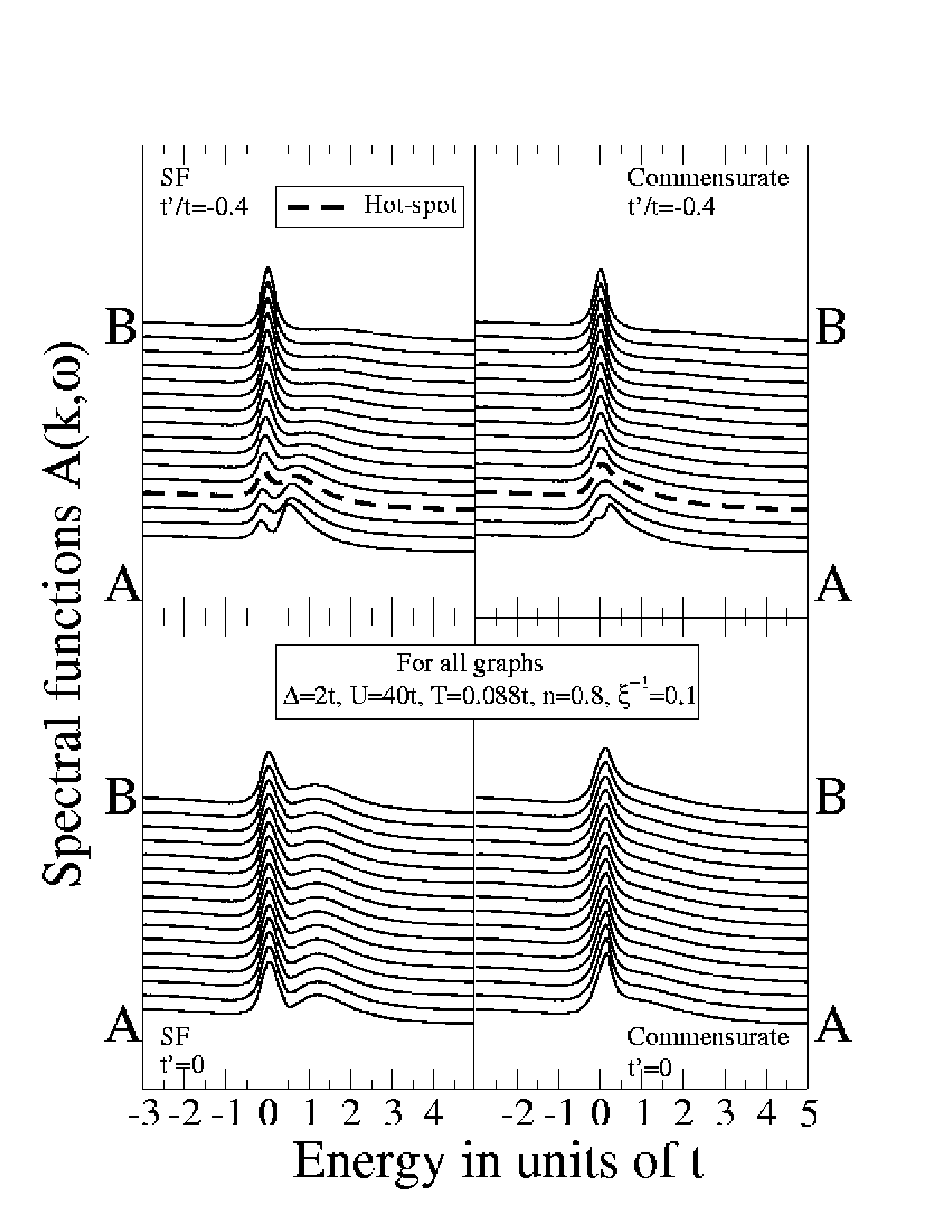}
\caption{Spectral functions $A({\bf k},\omega)$ 
obtained from the DMFT(NRG)+$\Sigma_{\bf k}$ calculations
for $U=40t$,
other parameters as in Fig.~\ref{sf_U4t_n08}.}
\label{sf_U40t_n08}
\end{figure}

For the case of a
doped Mott insulator ($U=40t$, $n=0.8$), the spectral functions obtained by the
DMFT+$\Sigma_{\bf k}$ approach are presented in Fig.~\ref{sf_U40t_n08}.
Qualitatively, the shapes of these spectral functions are similar to 
those shown on Fig.~\ref{sf_U4t_n08}. 
As was pointed out above, the strong Coulomb correlations lead to a
narrowing of the quasiparticle peak and a corresponding decrease of the  
pseudogap width. One should also note that in contrast to $U=4t$
the spectral functions are now less intensive, because part of the spectral weight is 
transferred to the upper Hubbard band located at about $40t$.

\begin{figure}[htb]
\includegraphics[clip=true,width=0.6\textwidth]{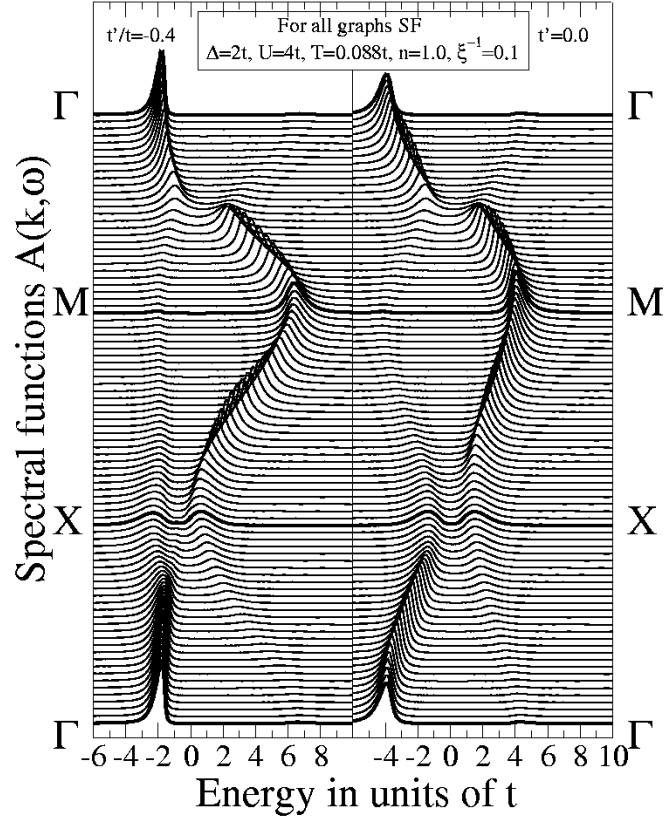}
\caption{Spectral functions $A({\bf k},\omega)$ obtained from the
DMFT(NRG)+$\Sigma_{\bf k}$ calculations
along high-symmetry directions of first Brilloin zone
$\Gamma(0,0)\!-\!\rm{X}(\pi,0)\!-\!\rm{M}(\pi,\pi)\!-\!\Gamma(0,0)$,
SF combinatorics (left row) and commensurate combinatorics (right column).
Other parameters are $U=4t$, $n=1.0$, $\Delta=2t$, $\xi^{-1}=0.1$ 
and temperature $T=0.088t$.
The Fermi level corresponds to zero.}
\label{n1_U4t_tri}
\end{figure}

\begin{figure}[htb]
\includegraphics[clip=true,width=0.6\textwidth]{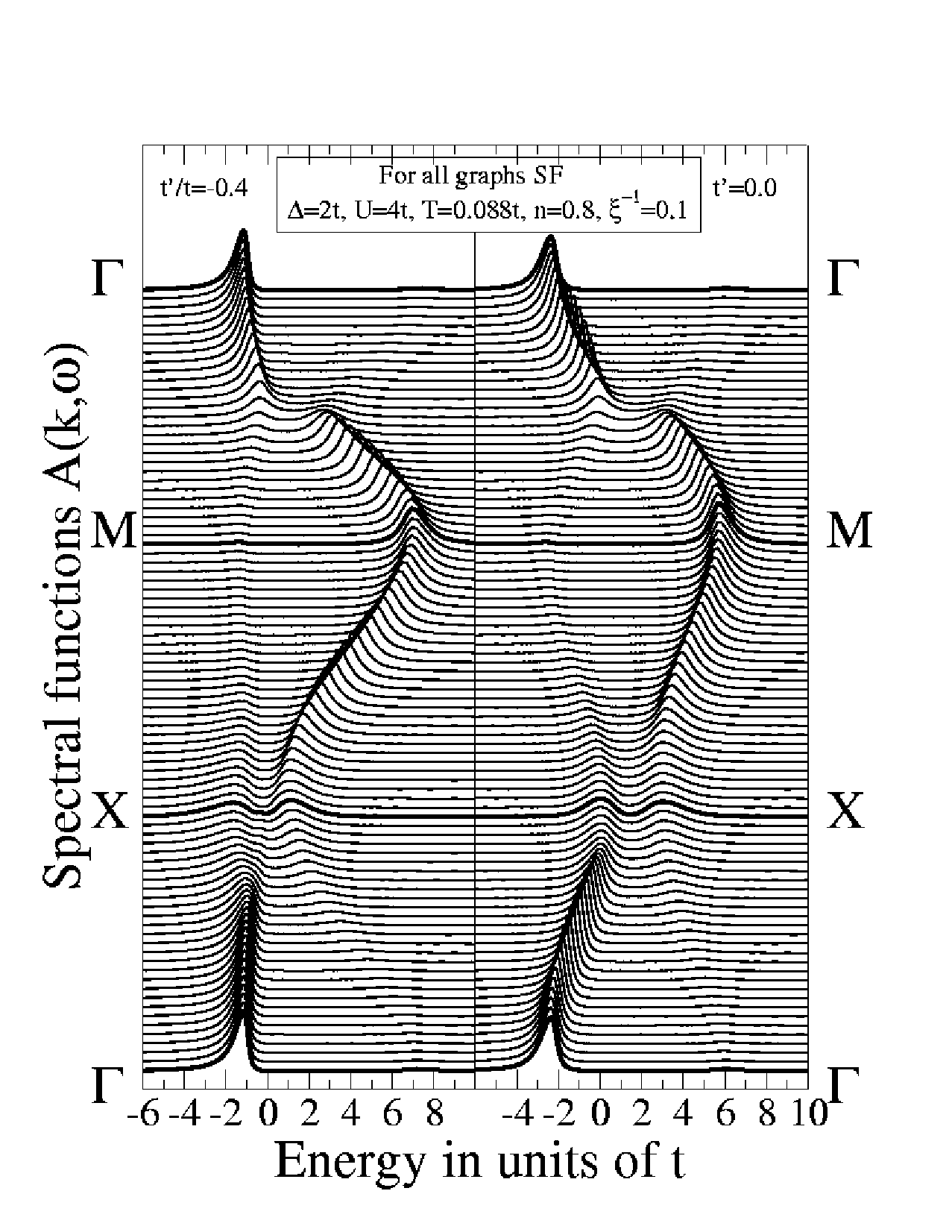}
\caption{Spectral functions $A({\bf k},\omega)$ obtained from the
DMFT(NRG)+$\Sigma_{\bf k}$ calculations along high-symmetry lines
in the $2D$ Brillouin zone for $n=0.8$,
other parameters as in Fig.~\ref{n1_U4t_tri}.}
\label{n08_U4t_tri}
\end{figure}

\begin{figure}
\includegraphics[clip=true,width=0.6\textwidth]{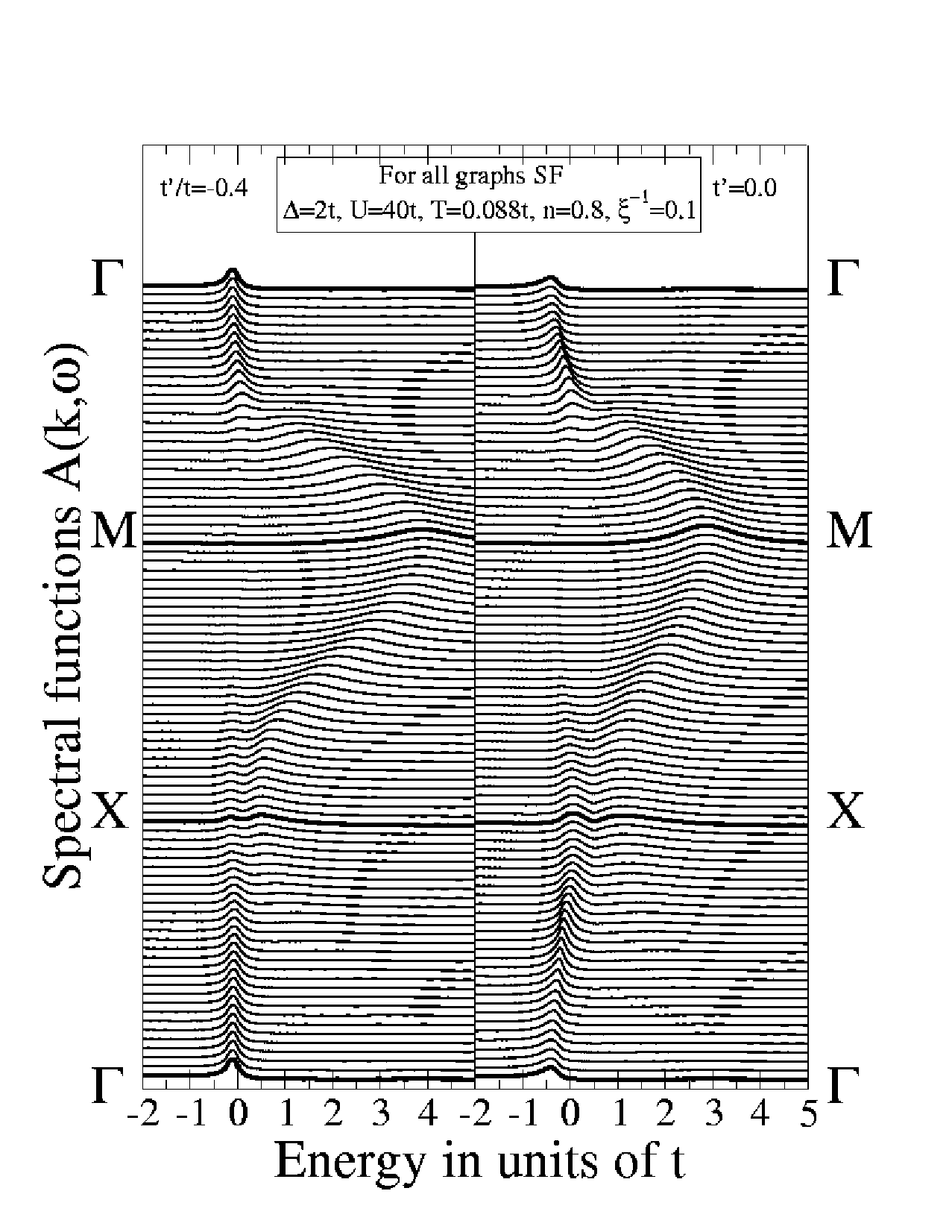}
\caption{Spectral functions $A({\bf k},\omega)$  obtained from the
DMFT(NRG)+$\Sigma_{\bf k}$ calculations along high-symmetry lines
in the $2D$ Brillouin zone for $U=40t$,
other parameters as in Fig.~\ref{n08_U4t_tri}.}
\label{tri_U40t_n08}
\end{figure}

Using another quite common choice of ${\bf k}$--points we can compute 
$A(\omega,{\bf k})$ along high--symmetry directions in the first 
Brillouin zone: $\Gamma(0,0)\!-\!\rm{X}(\pi,0)\!-\!\rm{M}(\pi,\pi)\!-\!\Gamma(0,0)$.  
The spectral functions for these 
${\bf k}$--points are colelcted in Figs.~\ref{n1_U4t_tri}, \ref{n08_U4t_tri} and
\ref{tri_U40t_n08} for the case of SDW--like fluctuations.
For all sets of parameters one can 
see a characteristic double peak pseudogap structure at the $X$ point and in 
the middle of $M\!-\!\Gamma$ direction.  A change of the filling leads mainly to a 
rigid shift of spectral functions with respect to the Fermi level. 

\begin{figure}
\includegraphics[clip=true,width=0.6\textwidth]{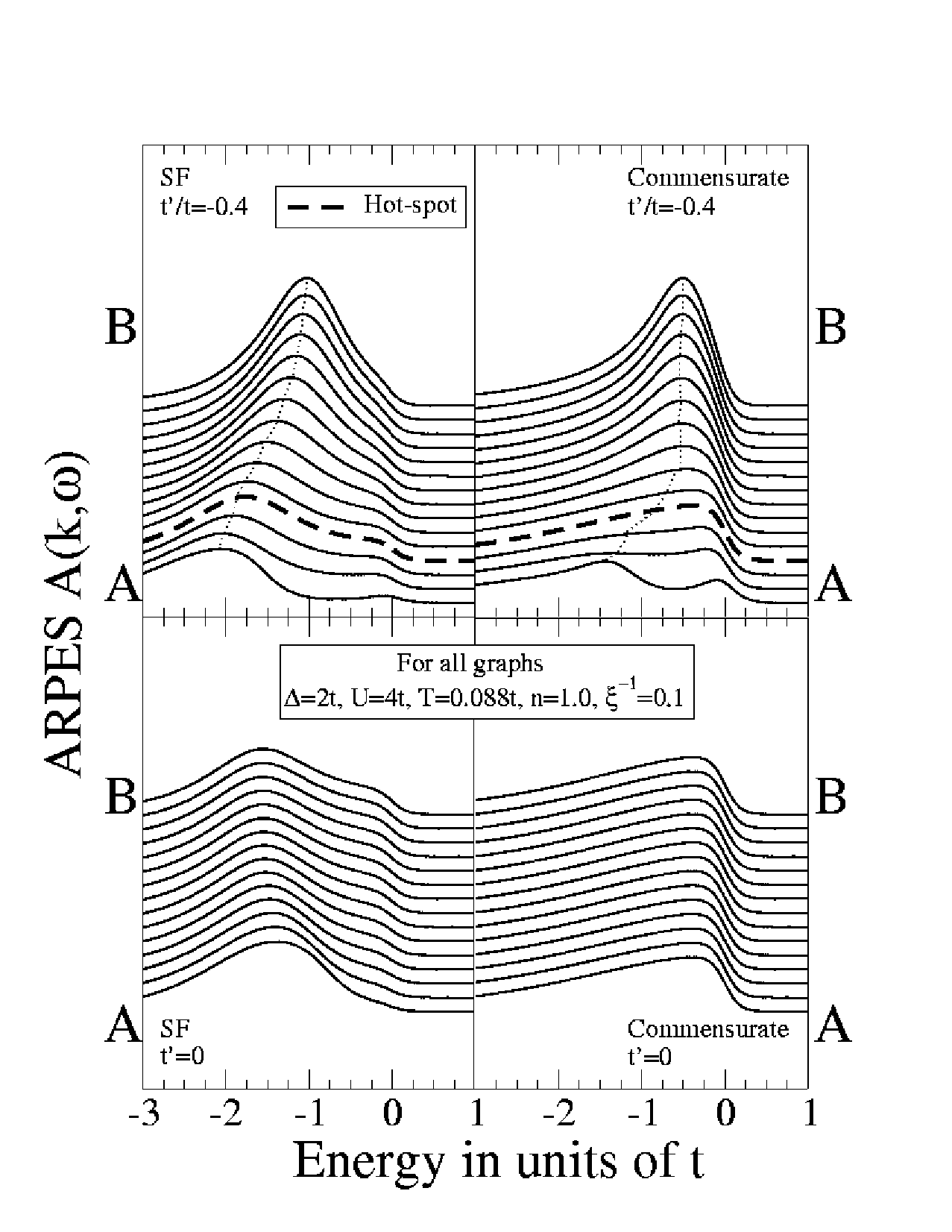}
\caption{ARPES  obtained from the
DMFT(NRG)+$\Sigma_{\bf k}$ calculations for $U=4t$ and $n=1.0$
along the lines in the first BZ as depicted by
Fig.~\ref{FS_shapes}, all other parameters as in Fig.~\ref{sf_U4t_n1}.}
\label{arpes_U4t_n1}
\end{figure}

\begin{figure}
\includegraphics[clip=true,width=0.6\textwidth]{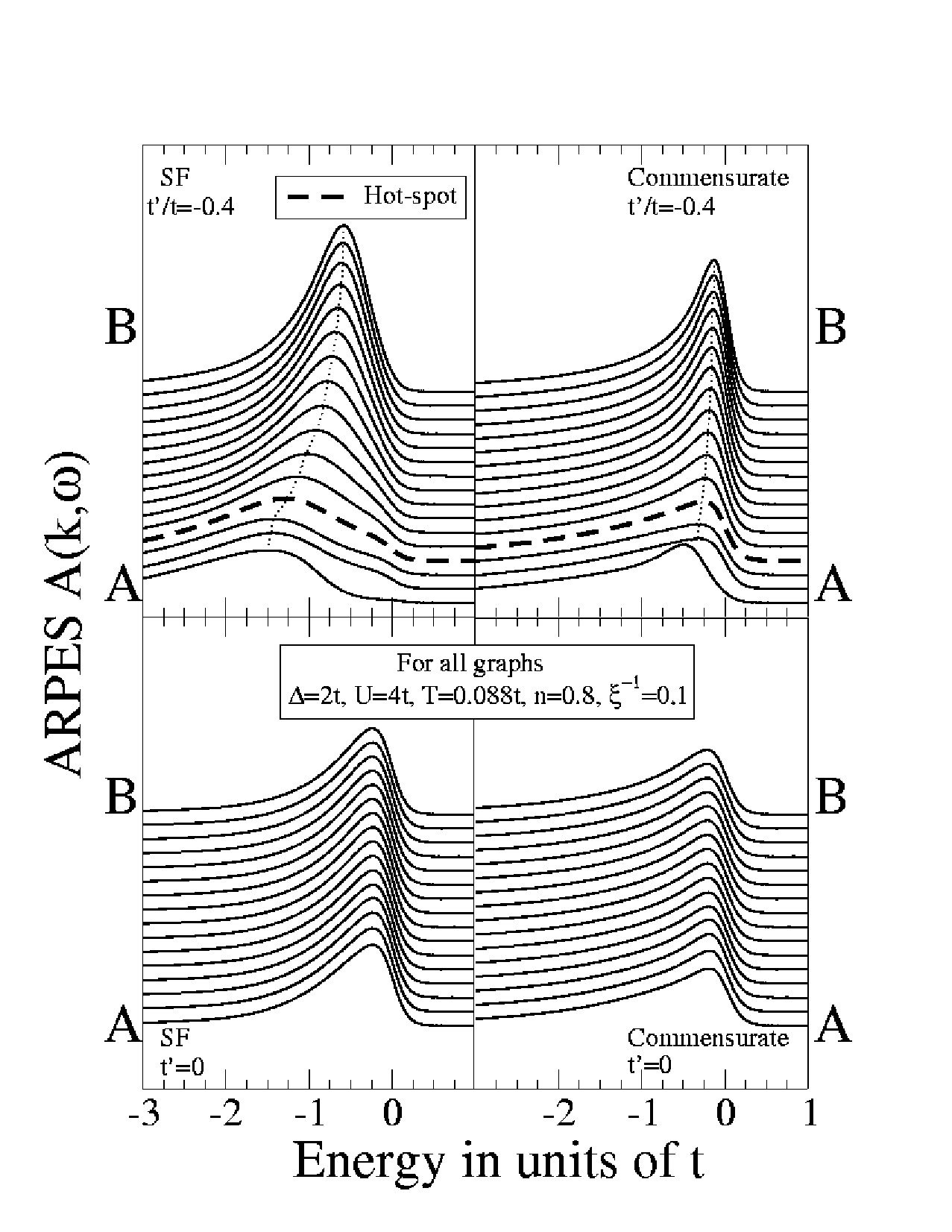}
\caption{ARPES  obtained from the
DMFT(NRG)+$\Sigma_{\bf k}$ calculations for $U=4t$ and $n=0.8$
along the lines in the first BZ as depicted by
Fig.~\ref{FS_shapes}, all other parameters as in Fig.~\ref{sf_U4t_n08}.}
\label{arpes_U4t_n08}
\end{figure}

\begin{figure}
\includegraphics[clip=true,width=0.6\textwidth]{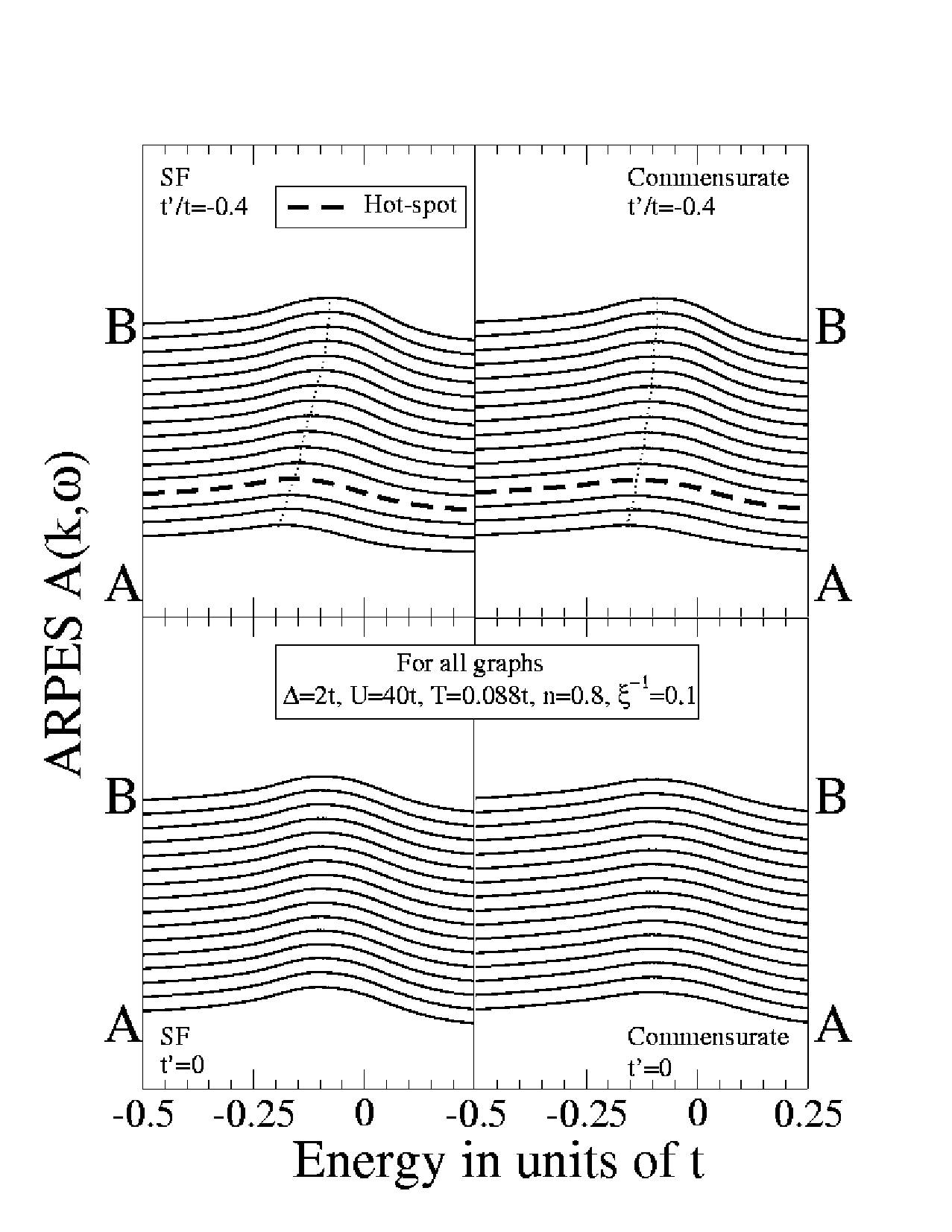}
\caption{ARPES  obtained from the
DMFT(NRG)+$\Sigma_{\bf k}$ calculations for $U=40t$ and $n=0.8$
along the lines in the first BZ as depicted by
Fig.~\ref{FS_shapes}, all other parameters as in Fig.~\ref{sf_U40t_n08}.}
\label{arpes_U40t_n08}
\end{figure}

With the spectral functions we are now of course in a position to calculate
angle resolved photoemission spectra (ARPES), which is the most direct
experimental way to observe pseudogap in real compounds. For that purpose, we only
need to multiply our results for the spectral functions with the Fermi function at
temperature $T=0.088t$. the resulting DMFT+$\Sigma_{\bf k}$ ARPES spectra are
presented in Figs.~\ref{arpes_U4t_n1}, \ref{arpes_U4t_n08} and \ref{arpes_U40t_n08}
One should note that for $t'/t=-0.4$ (upper panel of
Figs.~\ref{arpes_U4t_n1}, \ref{arpes_U4t_n08} and \ref{arpes_U40t_n08}) as
${\bf k}$ goes from point ``A'' to point ``B'' the
peak situated slightly below the Fermi level changes its
position and moves down in energy.
Simultaneously it becomes more broad and less intensive.
The dotted line guides the motion of the peak maximum.
Also at the ``hot--spot'' and further to point B one can see some signs
of the double--peak structure.
Such behaviour of the peak in the ARPES is rather reminiscent of those observed
experimentally in underdoped cuprates \cite{MS,Sch,Kam}.

\section{Conclusion}
\label{concl}

To summarize, we propose a generalized DMFT+$\Sigma_{\bf k}$
approach, which is meant to take into account the important
effects due to non--local correlations in a systematic, but to some extent
phenomenological fashion.
The main idea of this extension is to stay within a
usual effective Anderson impurity analogy, and introducing length scales due
to non-local correlation via the effective medium (``bath'') appearing in the standard
DMFT. This becomes possible by incorporating scattering processes of fermions
in the ``bath'' from non-local collective SDW--like antiferromagnetic spin (or
CDW--like charge) fluctuations.
Such a generalization of the DMFT allows one to overcome the well--known 
shortcoming of ${\bf k}$--independence of self--energy of standard DMFT.  
It in turn opens the possibility to access the physics of low--dimensional 
strongly correlated systems, where different types of spatial fluctuations 
(e.g. of some order parameter), become important in a non-perturbative way at
least with respect to the important local dynamical correlations. 
However, we must stress that our procedure in no way introduces any kind of
systematic $1/d$--expansion, being only a qualitative method to include
length scale into DMFT.

In our present study we addressed the problem of pseudogap formation in the
strongly correlated metallic state.  We showed evidence that the pseudogap appears at 
the Fermi level within the quasiparticle peak, introducing a new
small energy scale of the order of $\Delta$ in the DOSs and spectral functions
$A(\omega,{\bf k})$.

Let us stress, that our generalization of DMFT leads to non--trivial 
and in our opinion physically sensible ${\bf k}$--dependence of spectral functions. 
In contrast, in a recent work by Haule~{\it et al.}~\cite{HW}
the extended DMFT approach was used to demonstrate 
pseudogap formation in DOS due to dynamic Coulomb correlations only.
However, within the approach of Ref.~\cite{HW} there is no way to obtain
${\bf k}$--dependence of spectral functions beyond that originating from the
bare electronic energy dispersion which is actually observed in experiments.
Of course, similar results and observations were in recent years also made using
the cluster mean-field theories \cite{TMrmp}. 
The major advantage of our approach over these cluster mean-field theories is,
that we stay in an effective single-impurity picture. This means that 
our approach is computationally much less
expensive and therefore also easily generalizable to multi-orbital systems.

\section{Acknowledgements}

We are grateful to A. Kampf for useful discussions.
This work was supported in part by RFBR grants 05-02-16301 (MS,EK,IN)
RFFI-GFEN-03-02-39024\_a (VA,IN), RFFI-04-02-16096 (VA,IN), by the joint 
UrO-SO project $No.$ 22 (VA,IN), and programs of the
Presidium of the Russian Academy of Sciences (RAS) ``Quantum macrophysics''
and of the Division of Physical Sciences of the RAS ``Strongly correlated
electrons in semiconductors, metals, superconductors and magnetic
materials''. I.N. acknowledges support
from the Dynasty Foundation and International
Centre for Fundamental Physics in Moscow program for young
scientists 2005), Russian Science Support Foundation program for
young PhD of Russian Academy of Science 2005. One of us (TP) further acknowledges
supercomputer support from the Norddeutsche Verbund f\"ur Hoch- und
H\"ochstleistungsrechnen.

\newpage


\appendix

\section{Derivation of generalized DMFT+$\Sigma_{\bf k}$ approach}
\label{A}

In this appendix we present a derivation of the generalized DMFT+$\Sigma_{\bf k}$ scheme
for the Hubbard model
\begin{equation}
H=-\sum_{ij,\sigma}{t_{ij}c_{i\sigma}^\dagger c_{j\sigma}}+
U\sum_i{n_{i\uparrow}n_{i\downarrow}}
\label{extended_Hubbard_model},
\end{equation}
using a diagrammatic approach.
The single--particle Green function in Matsubara representation is as usual given by
\begin{equation}
G_{\bf k}(i\omega)=\frac{1}{i\omega+\mu-\varepsilon({\bf k})
-\Sigma(i\omega,{\bf k})}
\label{GkH}
\end{equation}
To establish the standard DMFT one invokes the limit of infinite dimensions $d \to \infty$.
In this limit only local contributions to electron self--energy survive \cite{vollha93,georges96}, i.e.\ 
$\Sigma_{ij} \to \delta_{ij} \Sigma_{ii}$ or, in reciprocal space,
$\Sigma(i\omega,{\bf k}) \to \Sigma(i\omega)$.

\begin{figure}
\includegraphics[clip=true,width=0.8\textwidth]{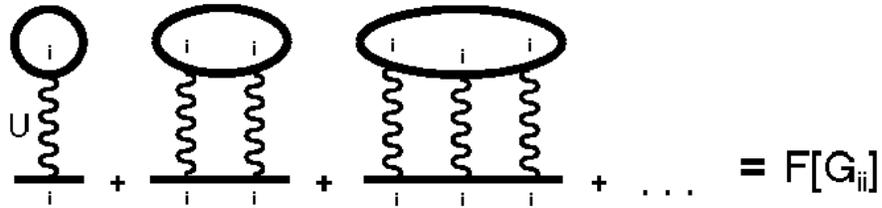}
\caption{Local ``skeleton'' diagrams for the DMFT self--energy $\Sigma$.
Wavy lines represent the local (Hubbard) Coulomb interaction $U$, full lines 
denote the local Green function $G_{ii}$.}
\label{sigmDMFT}
\end{figure}

In Fig. \ref{sigmDMFT} we show examples of ``skeleton'' diagrams for the local
self -- energy, contributing in the limit of $d \to \infty$. The complete series 
of these and similar diagrams defines the local self -- energy as a functional
of the local Green function
\begin{equation}
\Sigma = F[G_{ii}]\;,
\label{FGii}
\end{equation}
where
\begin{equation} 
G_{ii}(i\omega)=\frac{1}{N}\sum_{\bf k}\frac{1}{i\omega+\mu
-\varepsilon({\bf k})-\Sigma(i\omega)}.
\label{Gii}
\end{equation}
One then defines the ``Weiss field''
\begin{equation}
{\cal G}^{-1}_{0}(i\omega)=\Sigma(i\omega)+G^{-1}_{ii}(i\omega)
\label{Wssn}
\end{equation}
which is used to set up the effective impurity problem with an effective action
given by (\ref{Seff}). Via Dyson's equation the Green function (\ref{AndImp})
for this effective impurity problem can be written as
\begin{equation}
G_d(i\omega)=\frac{1}{{\cal G}^{-1}_{0}(i\omega)-\Sigma_d(i\omega)}
\label{Gd}
\end{equation}
and the ``skeleton'' diagrams for self--energy $\Sigma_d$ are just the same as
shown in Fig. \ref{sigmDMFT}, with the replacement $G_{ii} \to G_d$.
Thus we get
\begin{equation}
\Sigma_d = F[G_d],
\label{FGd}
\end{equation}
where $F$ is the same functional as in (\ref{FGii}). The two equations
(\ref{Gd}) and (\ref{FGd}) define both $G_d$ and $\Sigma_d$ for a given
``Weiss field'' ${\cal G}_0$. On the other hand, for the local
$\Sigma$ and $G_{ii}$ of the initial (Hubbard) problem we have precisely the
same  pair of equations, viz (\ref{FGii}) and (\ref{Wssn}), and ${\cal G}_0$ in both
problems is just the same, so that
\begin{equation}
\Sigma=\Sigma_d; \qquad G_{ii}=G_d.
\label{eqv}
\end{equation}
Thus, the task of finding the local self--energy of the 
$(d\to\infty)$ Hubbard model is eventually reduced to the calculation of the self--energy
of an effective quantum impurity problem defined by effective action of Eq.
(\ref{Seff}).

\begin{figure}
\includegraphics[clip=true,width=0.8\textwidth]{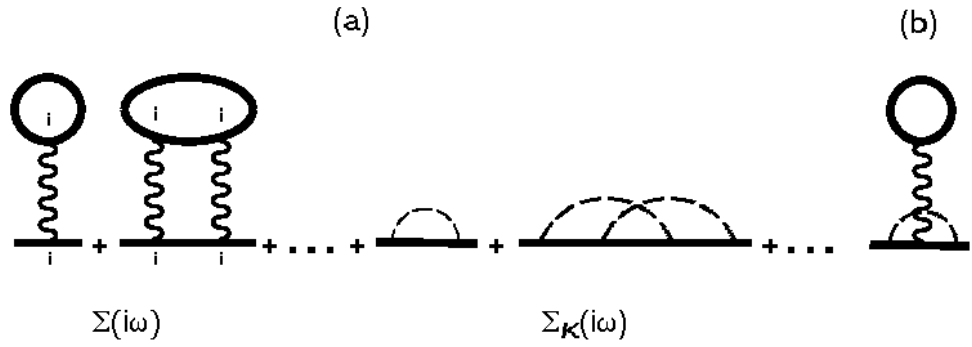}
\caption{Typical ``skeleton'' diagrams for the self--energy in the
DMFT+$\Sigma_{\bf k}$ approach.
The first two terms are DMFT self--energy diagrams; the
middle two diagrams show contributions to the
non-local part of the self--energy from
spin fluctuations (see section\protect\ref{kself}) represented
as dashed lines;
the last diagram (b) is an example of neglected diagramms leading to
interference
between the local and non-local parts. }
\label{dDMFT_PG}
\end{figure}

Consider now non -- local contribution to the self -- energy. If we neglect
interference between local and non--local contributions (as given e.g.
by the diagram shown in Fig.\ref{dDMFT_PG}(b)), the full self--energy is approximately
determined by the sum of these two contributions. ``Skeleton'' diagrams
for the non-local part of the self--energy, $\Sigma_{\bf k}(i\omega )$, 
are then those shown in Fig. \ref{dDMFT_PG}(a), where
the full line denotes the Green function $G_{\bf k}$ of Eq. (\ref{Gk}), 
while broken lines denote the interaction with static Gaussian spin (charge)
fluctuations.

The local contribution to the self--energy is again defined by the functional
(\ref{FGii}) via the local Green function $G_{ii}$, which is now given by
(\ref{Gloc}). Introducing again a ``Weiss field'' via (\ref{Wssn}) and repeating all
previous arguments, 
we again reduce the task of finding the local  part of the self--energy to
the solution of an ``impurity'' problem with an effective action (\ref{Seff}).

To determine the non--local contribution $\Sigma_{\bf k}(i\omega )$ we first
introduce
\begin{equation} 
{\cal G}_{0\bf k}(i\omega)=\frac{1}{G_{\bf k}^{-1}(i\omega)+
\Sigma_{\bf k}(i\omega)}
=\frac{1}{i\omega+\mu-\varepsilon({\bf k})-\Sigma(i\omega)}
\label{G0pg}
\end{equation}
as the ``bare'' Green function for electron scattering by static Gaussian
spin (charge) fluctuations. The assumed static nature of these fluctuations
allows to use the method of Refs.\cite{MS79,Sch,KS} and the calculation of the 
non--local part of the self--energy $\Sigma_{\bf k}(i\omega )$ reduces to the recursion
procedure defined by Eqs.~(\ref{Sk}) and (\ref{rec}). The choice of the ``bare''
Green function Eq.~(\ref{G0pg}) guarantees that the Green function ``dressed'' 
by fluctuations
$G_{\bf k}^{-1}(i\omega)={\cal G}_{0\bf k}^{-1}(i\omega)-\Sigma_{\bf k}(i\omega)$, 
which enters into the ``skeleton'' diagrams for $\Sigma_{\bf k}(i\omega)$, just
coincides with the full Green functions $G_{\bf k}(i\omega)$.

Thus we obtain a fully self--consistent scheme to calculate both local
(due to strong single--site correlations) and non--local (due to short--range
fluctuations) contributions to electron self--energy.

\section{$\Delta$ in the Hubbard model.}
\label{B}

In this Appendix we derive the explicit microscopic expression for
pseudogap amplitude $\Delta$ given in (\ref{DeltHubb}). Within the two--particle
self--consistent approach of Ref.~\cite{VT}, valid for medium values of
$U$,  and neglecting charge fluctuations, we can write down an expression for the
electron self--energy of the form used in (\ref{Gk}), with
\begin{equation}
\Sigma_{\sigma}(i\omega)=Un_{-\sigma}
\label{Sigloc}
\end{equation}
as the lowest order local contribution due to the on--site Hubbard interaction,
surviving in the limit of $d\to\infty$, and exactly accounted for in DMFT
(with all higher--order contributions).
Non--local contribution to the self--energy (vanishing for $d\to\infty$)
due to interaction with spin--fluctuations then lead to the expression
\begin{equation}
\Sigma_{\vec k}(i\omega)=\frac{U}{4}\frac{T}{N}\sum_m\sum_{\bf q}
U_{sp}\chi_{sp}({\bf q},\nu_m)G_0({{\bf{k+q}},i\omega+i\nu_m})\;,
\label{Signloc}
\end{equation}
where
\begin{equation}
U_{sp}=g_{\uparrow\downarrow}(0)U,\qquad
g_{\uparrow\downarrow}(0)=\frac{<n_{i\uparrow}n_{i\downarrow}>}
{<n_{i\uparrow}><n_{i\downarrow}>}
\label{Usp}
\end{equation}
with $<n^2_{\sigma}>=<n_{\sigma}>$ and
$<n_{i\uparrow}>=<n_{i\downarrow}>=\frac{1}{2}n$ in the paramagnetic phase.
For the dynamic spin susceptibility $\chi_{sp}({\bf q},\nu_m)$ we use
the standard Ornstein--Zernike form \cite{VT}, similar to that
used in spin--fermion model \cite{Sch}, which describes enhanced
scattering with momenta transfer close to antiferromagnetic vector
${\bf Q}=(\pi/a,\pi/a)$. With these approximations, we can write down the following
expression for the non--local contribution to the self--energy \cite{Sch,KS}:
\begin{eqnarray}
\Sigma_{\vec k}(i\omega)=\frac{1}{4}UU_{sp}\frac{T}{N}\sum_m\sum_{\vec q}
\chi_{sp}({\bf q},\nu_m)\frac{1}{i\omega+i\nu_m+\mu-\varepsilon({\bf k+q})}
\approx\nonumber\\
\approx\frac{1}{4}UU_{sp}\frac{T}{N}\sum_m\sum_{\vec q}\chi_{sp}({\bf q},\nu_m)
\sum_{\vec q}S({\vec q})\frac{1}{i\omega+\mu-\varepsilon({\bf k+q})}
\equiv\nonumber\\
\equiv\Delta^2\sum_{\vec q}S({\bf q})\frac{1}{
i\omega+\mu-\varepsilon({\bf k+q})}=\nonumber\\
=\frac{\Delta^2}{i\omega+\mu-\varepsilon({\bf p+Q})
+i(|v^x_{\bf p+Q}|+|v^y_{\bf p+Q}|)\kappa {\rm sign}\omega}.
\label{SigKS}
\end{eqnarray}
Here we have introduced the static form factor \cite{KS}
\begin{equation}
S({\vec q})=\frac{2\xi^{-1}}{(q_x-Q_x)^2+\xi^{-2}}
\frac{2\xi^{-1}}{(q_y-Q_y)^2+\xi^{-2}}
\label{Sq}
\end{equation}
and the squared pseudogap amplitude
\begin{eqnarray}
\Delta^2=\frac{1}{4}UU_{sp}\frac{T}{N}
\sum_m\sum_{\vec q}\chi_{sp}({\bf q},\nu_m)=\nonumber\\
=\frac{1}{4}UU_{sp}[<n_{i\uparrow}>+<n_{i\downarrow}>
-2<n_{i\uparrow}n_{i\downarrow}>]=\nonumber\\
=\frac{1}{4}UU_{sp}\frac{1}{3}<{\vec S}_i^2>,
\label{Delta2}
\end{eqnarray}
where we have used the exact sum--rule for the susceptibility \cite{Sch,VT}.
Taking into account (\ref{Usp}) we immediately obtain (\ref{DeltHubb}).

Actually, the approximations made in (\ref{SigKS}) and (\ref{Sq}) allow for an exact
summation of the whole Feynman series for electron interaction with spin--fluctuations,
replaced by the static Gaussian random field.
Thus generalizing the one--loop approximation (\ref{SigKS}) eventually leads to the
basic recursion procedure given in (\ref{Sk}), (\ref{rec}) Refs.~\cite{Sch,KS}.

\begin{figure}
\includegraphics[clip=true,width=0.6\textwidth]{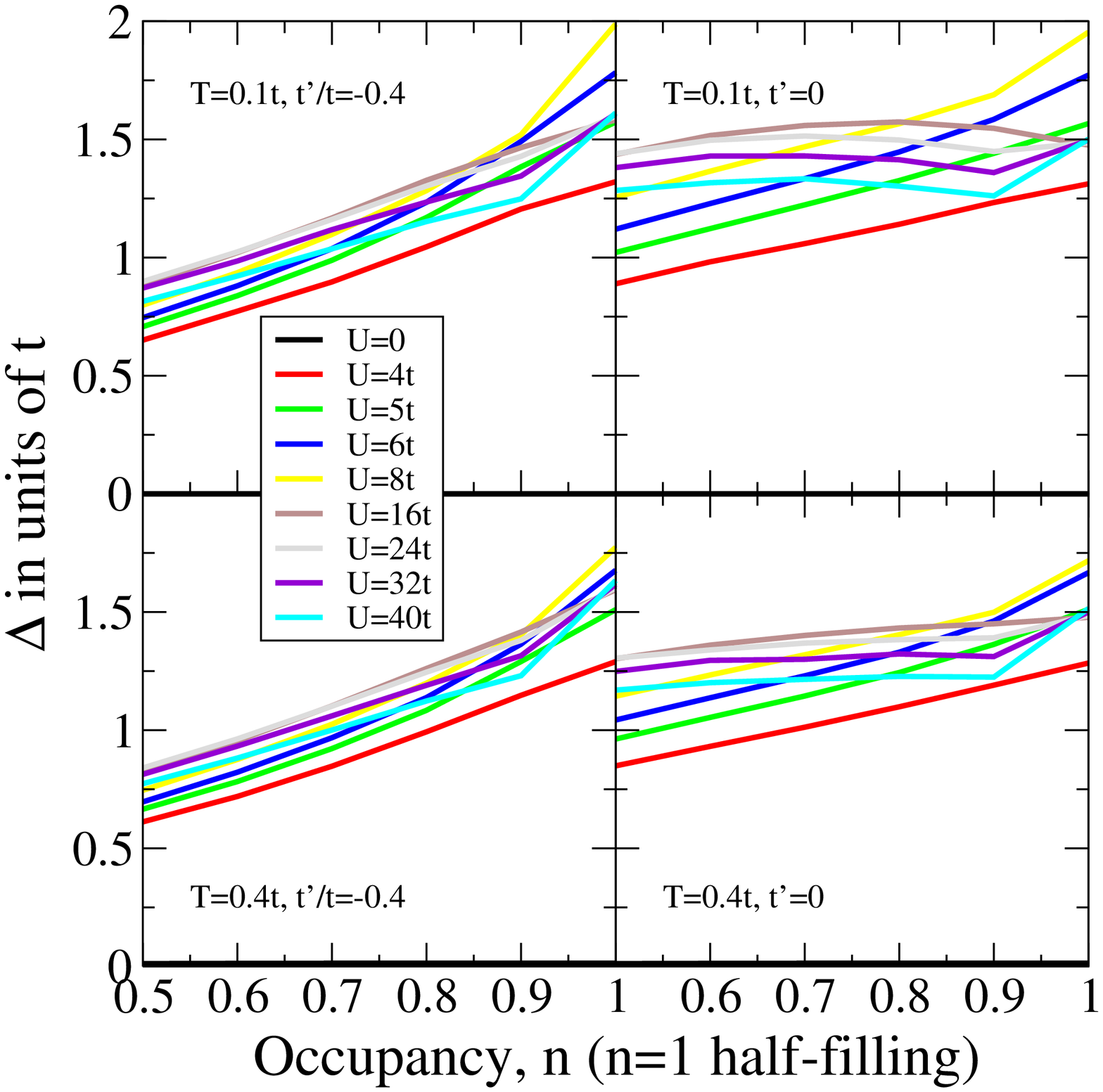}
\caption{Filling dependence of the pseudo gap potential $\Delta$
calculated with DMFT(QMC)  
for varying Coulomb interaction ($U$) and temperature ($T$)
on a two--dimensional square lattice with two sets of ($t,t'$).}
\label{delta}
\end{figure}

Using the DMFT(QMC) approach we computed occupancies $<n_{i\uparrow}>$, $<n_{i\downarrow}>$
and double occupancies $<n_{i\uparrow}n_{i\downarrow}>$ required to
calculate the pseudogap amplitude $\Delta$ of Eq. (\ref{Delta2})
In Fig. \ref{delta} the corresponding values of $\Delta$ are presented.
One can see that $\Delta$ grows when the filling goes to $n=1$.
While $U$ approaches $8t$ (the value of the bandwidth for a square lattice)
$\Delta$ as a function of $n$ grows monotonically. When $U$ becomes larger
than $W=8t$ (when a metal--insulator transition occurs) 
one can see a local minimum for 
$n=0.9$, which becomes more pronounced with further increase of $U$.
For $t'/t=-0.4$ and both temperatures the scatter of $\Delta$ values is 
smaller than for the case of $t'=0$. Also $\Delta$ has a rather weak 
temperature dependence.  
All values of $\Delta$ lie in the interval $\sim 
0.75t \div 2t$.  Therefore, for our computations we took only two  
characteristic values of $\Delta=t$ and $\Delta=2t$.

\pagestyle{empty}

\newpage

\newpage

\newpage

\newpage

\newpage

\newpage

\newpage

\newpage

\newpage

\newpage

\newpage

\newpage

\newpage

\newpage

\newpage

\newpage

\newpage

\begin{thebibliography}{99}
\bibitem{VHS} We have checked that with increasing
of Coulomb repulsion the Van--Hove singularity transforms gradiently into
quasiparticle peak for $U=(6\div 8)t$.
\end{thebibliography}
\end{document}